\documentstyle[aps,preprint]{revtex}

\begin{document}

\preprint{IFA-97/13}

\title{The extraction of hadronic parameters from experiments on
pionium}
\author{A. Gashi and G. Rasche}
\address{Institut f\"{u}r Theoretische Physik der Universit\"{a}t,
Winterthurerstrasse 190,\\ CH-8057 Z\"{u}rich, Switzerland}
\author{G.C. Oades}
\address{Institute of Physics and Astronomy, Aarhus University,
DK-8000 Aarhus C, Denmark}
\author{W.S. Woolcock}
\address{Department of Theoretical Physics, IAS,
The Australian National University, Canberra,\\ ACT 0200, Australia}
\date{submitted to Nuclear Physics A}
\maketitle
\begin{abstract}
We show how experimental values of the lifetime of the $1s$ level of
pionium and of the difference between the energies of the $2s$ and $2p$
levels yield values of the elements $a_{oc}$ and $a_{cc}$ respectively of
the $s$-wave scattering matrix for the two-channel $(\pi ^{+}\pi ^{-},\pi
^{0}\pi ^{0})$ system at the $\pi ^{+}\pi ^{-}$ threshold. We then develop
a method, using energy independent hadronic potentials which reproduce the
best available pion-pion phase shifts up to 500 MeV total energy in the
c.m. frame, for obtaining the values of the isospin invariant quantities $%
a_{0}^{2}-a_{0}^{0}$ and $2a_{0}^{0}+a_{0}^{2}$ from $a_{oc}$ and $a_{cc}$
respectively. We emphasise that the isospin invariant scattering lengths $%
a_{0}^{0}$ and $a_{0}^{2}$ universally used in the literature cannot be
considered to be purely hadronic quantities. 
\end{abstract}
\section{Introduction}
The $\pi^{+}\pi^{-}$ atom (pionium) has become of considerable experimental
and theoretical interest. Such atoms have been clearly observed at Dubna 
\cite{1} and a lower limit obtained for the lifetime $\tau_{1s}$ of the
lowest (1$s$) level \cite{2}. A large collaboration is engaged in an
experiment at CERN to measure $\tau_{1s}$ with an accuracy of about 10$\%$
\cite{3}. In the proposal for this experiment the possibility of measuring 
the difference $\Delta W=W_{2s}-W_{2p}$ between the energies of the 2$s$
and 2$p$ levels is also considered. The method was proposed by Nemenov
\cite{4} and its feasibility can be tested using the same experimental
setup as for the measurement of $\tau_{1s}$. 

A complementary approach has been suggested in Ref. \cite{5} by a group
at IUCF. They have considered the possibility of measuring the ratio of the
rates for the decay of $\pi^{+}\pi^{-}$, either in $ns$ bound states or
low-lying continuum states, into $\pi^{0}\pi^{0}$ and $\gamma\gamma$. The
known rate for the almost purely electromagnetic $\gamma\gamma$ decay would
then yield the rate for the $\pi^{0}\pi^{0}$ decay. Ref. \cite{5} indicates
however that a first search for a suitable reaction with $\pi^{+}\pi^{-}$
pairs in the final state has not yielded promising results. We therefore
concentrate in this paper on the results expected to come from the CERN
experiment. 

A measurement of $\tau_{1s}$ immediately yields the width of the 1$s$
level via the relation 
\begin{equation}
\Gamma_{1s}^{tot}({\rm eV})=0.658\;212\;2[\tau_{1s}({\rm fs})]^{-1}.
\label{eq1}
\end{equation}
The partial width for decay to $\gamma\gamma$ is very small, and can be
derived from QED for bosons \cite{6}. The result is 
\begin{equation}
\Gamma_{1s}^{\gamma\gamma}=\frac{1}{4}\alpha^{5}\mu_{c}=7.2 \times 10^{-4}
{\rm eV},  \label{eq2}
\end{equation}
$\alpha$ being the fine structure constant and $\mu_{c}$ the mass of a
charged pion. Subtracting Eq. (\ref{eq2}) from Eq. (\ref{eq1}) gives the
value of $\Gamma_{1s}^{\pi^{0}\pi^{0}}$, which we shall henceforth
abbreviate to $\Gamma$. Since $\Gamma$ is expected to be about 0.177 eV
and will be measured with an accuracy of around 10$\%$, no refined
corrections to the result in Eq. (\ref{eq2}) are needed. 

We expect then that values of $\Gamma$ and, later, $\Delta W$ will
become available from experiments on pionium. In Section 2 we shall go
briefly through the formalism for pionium, which is identical with that
given by Rasche and Woolcock \cite{7} for pionic hydrogen. We shall give
the explicit formulae for $a_{oc}$(fm) and $a_{cc}$(fm) in terms of
$\Gamma({\rm eV})$ and $\Delta$W(eV) respectively. The matrix 
\begin{equation}
{\bf {a} = \left( 
\begin{array}{cc}
a_{cc} & a_{oc} \\ 
a_{oc} & a_{oo}
\end{array}
\right)}  \label{eq3}
\end{equation}
is the $s$-wave scattering matrix for the two-channel $(\pi^{+}\pi^{-},
\pi^{0}\pi^{0})$ system at the $\pi^{+}\pi^{-}$ threshold. The subscript
$c$ refers to the $\pi^{+}\pi^{-}$ channel, $o$ to the $\pi^{0}\pi^{0}$
channel. 

The challenge we are addressing is to connect these numerical
results, derivable in principle from experiments on pionium, with our
knowledge of pion-pion scattering at low energies (by which we shall mean
that the total energy $W$ in the c.m. frame is less than about 500 MeV).
This knowledge comes from two sources, experiments in which pion-pion pairs
in the final states of various reactions are studied, and chiral
perturbation theory ($\chi$PT), which gives a low-energy representation of
the $\pi\pi$ scattering amplitudes based on QCD. The difficulty of the
problem lies in the subtle interplay of the two interactions present, the
hadronic and the electromagnetic. The values of $a_{oc}$ and $a_{cc}$ which
come from pionium experiments are not due to the hadronic interaction
alone: they are affected by the electromagnetic interaction. On the other
hand, $\chi$PT as it has been developed so far, with the quark mass
difference neglected and only the QCD lagrangian used, is a hadronic
interaction theory which is charge independent (isospin invariant).
Moreover, experiments above the $\pi^{+}\pi^{-}$ threshold are analysed to
give what are quoted as phase shifts for states of definite isospin
($\delta_{0}^{0}$ and $\delta_{0}^{2}$ for $\ell$=0, $I$=0 and 2;
$\delta_{1}^{1}$ for $\ell$=1, $I$ =1;$\cdots$). In spite of this, as we
shall see, neither $\chi$PT as it is presently applied, nor the analysis
of experiments, completely disentangles the electromagnetic interaction
from the hadronic. Nevertheless it is clear that electromagnetic
corrections need to be made, to relate the values of
$a_{oc}$ and $a_{cc}$ obtained from pionium to the values of the isospin
invariant $s$-wave scattering lengths $a_{0}^{0}$ and $a_{0}^{2}$ obtained
from above threshold experiments and $\chi$PT. 

One approach has been made in a recent preprint by Maltman and Wolf
\cite{8}, who attempt to include the electromagnetic interaction as an
additional symmetry breaking term in $\chi$PT in order to calculate the
corrections to the purely hadronic $\pi\pi$ scattering lengths. At this
stage of their work too few details are given to fully evaluate its
consequences for pionium. In this paper we use a potential model,
considered previously by Moor, Rasche and Woolcock \cite{9}, to calculate
the electromagnetic corrections described above. 

Moor et al. \cite{9} attempted to calculate $a_{oc}-\frac{\sqrt{2}}{3}
\left(a_{0}^{2}-a_{0}^{0}\right)$ by distinguishing between the physical
situation and the purely hadronic situation in which the electromagnetic
interaction is switched off $(\alpha=0)$. In the first, the charged and
neutral pions have their physical masses $\mu_{c}$ and $\mu_{o}$
respectively and the hadronic and electromagnetic interactions are present.
In the second, there is only the hadronic interaction and, in accordance
with current opinion \cite{10,11,12}, it is assumed that all the pions have
almost the same mass, which is extremely close to $\mu_{o}$ , and that
there is practically no isospin breaking of the hadronic scattering
amplitudes. Ref. \cite{9} connects these two situations by means of a
potential model. For the purely hadronic situation, all the pions were
taken to have the mass $\mu_{o}$ and only a charge independent hadronic
potential was present. Simple square well potentials were used in a
relativistic Schr\"{o}dinger equation in order to reproduce at the energy
$W=2\mu_{o}$ the purely hadronic scattering lengths $a_{0}^{I}, I=0, 2$,
obtained from $\chi$PT. For the physical situation a two-channel $s$-wave
relativistic Schr\"{o}dinger equation was used, with physical masses and
(relativistic) c.m. momenta at a fixed energy, the potential matrix being
the sum of a hadronic potential and a Coulomb potential for extended
charge distributions. In the spirit of $\chi$PT it was assumed that the
mass difference is the only source of symmetry breaking, so that the
hadronic potential for the physical situation was taken to be exactly the
same as the charge independent potential for the purely hadronic situation.
The two-channel equation for the physical situation was then solved to
obtain $a_{oc}$ at $W=2\mu_{c}$, and thus the difference
$a_{oc}-\frac{\sqrt{2}}{3}(a_{0}^{2}-a_{0}^{0})$. 

There are two serious problems with the work in Ref. \cite{9}. First, the
purely hadronic scattering lengths $a_{0}^{0}$ and $a_{0}^{2}$ were
obtained by  inserting the mass $\mu_{o}$ into the $\chi$PT equations of
Gasser and Leutwyler \cite{10}. However some constants in those equations
are fixed by using experimental data extrapolated to the $\pi^{+}\pi^{-}$
threshold $W=2\mu_{c}$. This is an inconsistent procedure. In the present
state of our knowledge it is not possible to obtain purely hadronic
scattering lengths at the purely hadronic threshold $W=2\mu_{o}$. One is
prevented by the fact that, in the analysis of experiments and in the
practical use of $\chi$PT, the $\pi\pi$ system at the hadronic level is
treated as an isospin invariant system with all the pions having the mass
$\mu_{c}$. On the basis of this assumption charge independent phase shifts 
are obtained from experiments. This information is then used, in
conjunction with the $\chi$PT equations (the pion mass being taken as
$\mu_{c}$), to generate $s$-wave phase shifts $\delta_{0}^{0}$ and
$\delta_{0}^{2}$ for $W\geq 2\mu_{c}$.  Sets of these
phases have been provided to us by Gasser \cite{13}. We note here that
their behaviour at low energies is different from that in the fits of
Froggatt and Petersen \cite{14} and Lohse et al. \cite{15}. Using the
expansion 
\begin{equation}
\tan\delta_{0}^{I}/q_{c}=a_{0}^{I}+\alpha_{0}^{I}q_{c} ^{2}+\cdots, I=0, 2,
\label{eq4}
\end{equation}
the $\chi$PT equations show clearly that $\alpha_{0}^{0}>0, \alpha_{0}^{2}
< 0$, so that the curvatures $\alpha_{0}^{I}$ have the same signs as the
respective scattering lengths $a_{0}^{I}$. For this reason, we have
reservations about the potentials constructed by Sander, Kuhrts and von
Geramb \cite{16}, based on the fits of Refs. \cite{14,15}. 

The second problem in the work of Ref. \cite{9} is that simple square
well potentials were used. To obtain the values of both $a_{0}^{I}$ and
$\alpha_{0}^{I}$ given by $\chi$PT, it is necessary for such potentials to
be energy dependent. Further, Moor et al. calculated the difference
between the purely hadronic scattering matrix at $W=2\mu_{o}$ and the
physical scattering matrix at $W=2\mu_{c}$ without taking into account
the variation of their potentials with energy. This particular problem no
longer arises in the present work, where we shall calculate electromagnetic
corrections at $W=2\mu_{c}$. Nevertheless, there remain strong reasons
for desiring an energy independent hadronic potential matrix 
\begin{equation}
{\bf {V}^{had}=\left(
\begin{array}{cc}
\frac{2}{3}V_{0}^{0}+\frac{1}{3}V_{0}^{2} & \frac{\sqrt{2}}{3}
(V_{0}^{2}-V_{0}^{0}) \\ 
\frac{\sqrt{2}}{3}(V_{0}^{2}-V_{0}^{0}) & \frac{1}{3}V_{0}^{0}+\frac{2}{3}
V_{0}^{2}
\end{array}
\right).}  \label{eq5}
\end{equation}
If such potentials $V_{0}^{I}$ can be found, the representation of the
strong interaction by a potential matrix which remains unchanged when the
electromagnetic interaction is switched on is placed on much firmer ground.
The assumption that the $\pi\pi$ system at the hadronic level can be
treated as an isospin invariant system with all the pions having the mass 
$\mu_{c}$ then translates into a two-channel Schr\"{o}dinger equation for
which the equations decouple in the isospin basis. The mass $\mu_{c}$
plays the role of a convenient reference mass and we emphasise that the
scattering lengths $a_{0}^{I}$ obtained from the potentials $V_{0}^{I}$
by using a Schr\"{o}dinger equation with this mass are isospin invariant
but not purely hadronic quantities. 

There is a further question about how to construct energy
independent potentials $V_{0}^{I}$. If the phase shifts $\delta_{0}^{I}$
were perfectly known and the process remained elastic at all energies, it
would be possible by inversion to construct potentials $V_{0}^{I}$.
However, there are substantial uncertainties in the values of
$\delta_{0}^{I}$ (and electromagnetic corrections in the sense of
Ref. \cite{17} have not been made), there are inelastic processes at
sufficiently high energy and, most important of all, there is no reason
to believe that the hadronic interaction can be represented by potentials
except at low energies. We therefore do not use the results of
Ref. \cite{16} but instead find potentials of simple shape, characterised
by a small number of parameters, which can be fixed so that the potentials 
reproduce the phase shifts well up to $W=500$ MeV. In fact double square
wells with in all four parameters work quite satisfactorily. A sum of two
gaussians or of two lorentzian shapes were also tried, but the fit to the
phase shifts could not be further improved and it was more difficult to
ensure that the potentials had no bound states. 

The double square well energy independent potentials $V_{0}^{I}$ are
fed into the one-channel Schr\"{o}dinger equations 
\begin{equation}
\left(\frac{d^{2}}{dr^{2}}+q_{c}^{2}-\mu_{c}V_{0}^{I}(r)\right)u(r)=0, I=0,
2,  \label{eq6}
\end{equation}
and the ranges and depths of the two wells adjusted to obtain the best
possible fit to the phase shifts $\delta_{0}^{I}$ up to $W=500$ MeV.
Eq. (\ref{eq6})
also yields the scattering lengths $a_{0}^{I}$ and we denote by ${\bf {
\tilde{a}}}$ the matrix 
\begin{equation}
{\bf {\tilde{a}}=\left(
\begin{array}{cc}
\frac{2}{3}a_{0}^{0}+\frac{1}{3}a_{0}^{2} & \frac{\sqrt{2}}{3}
(a_{0}^{2}-a_{0}^{0}) \\ 
\frac{\sqrt{2}}{3}(a_{0}^{2}-a_{0}^{0}) & \frac{1}{3}a_{0}^{0}+\frac{2}{3}
a_{0}^{2}
\end{array}
\right).}  \label{eq7}
\end{equation}

For the physical situation we have the two-channel Schr\"{o}dinger
equation with physical masses 
\begin{equation}
\left({\bf 1}_{2}\frac{d^{2}}{dr^{2}}+{\bf Q}^{2}-2{{\bf MV}^{ had}}(r)
-2{{\bf MV}^{C}} (r)\right){\bf u}(r)=0,  \label{eq8}
\end{equation}
where ${\bf {V}^{had}}$ is given by Eq. (\ref{eq5}) and 
\[
{\bf Q}=\left(
\begin{array}{cc}
q_{c} & 0 \\ 
0 & q_{o}
\end{array}
\right), 2{\bf M}=\left(
\begin{array}{cc}
\mu_{c} & 0 \\ 
0 & \mu_{o}
\end{array}
\right), {{\bf V}^{C}}(r)=\left(
\begin{array}{cc}
V^{C}(r) & 0 \\ 
0 & 0
\end{array}
\right). 
\]
The relativistic c.m. momenta $q_{c},q_{o}$ in the two channels are
calculated from the same value of $W$; $V^{C}(r)$ is the Coulomb potential
between extended charge distributions. The coupled equations (\ref{eq8}) are
integrated numerically to give the scattering matrix ${\bf {K}}$ as a
function of $W$. The details are exactly the same as those given in Section
3 of Ref. \cite{18} for the coupled $(\pi^{-}p, \pi^{0}n)$ system; the
matrix ${\bf {A}}$ introduced in Eq. (65) of that paper we now denote by
${\bf {K}}$, in accordance with the altered notation of Ref. \cite{7}.
Extrapolation to $W=2\mu_{c}$ gives the threshold scattering matrix
${\bf {a}}={\bf {K}} \left(2\mu_{c}\right)$ of Eq. (\ref{eq3}) and thus the
difference ${\bf {\tilde{a}}-{a}}$, which is the final goal of our
calculation. The values of $a_{oc}$ and $a_{cc}$ obtained from experiments
on pionium can then be corrected to give values of $(a_{0}^{2}-a_{0}^{0})$
and $(2a_{0}^{0}+a_{0}^{2})$ which can be compared with the values
obtained from above threshold experiments and $\chi$
PT. In Section 3 we shall give further details of the calculations sketched
above, and the numerical results for the electromagnetic corrections. 

\section{Determination of scattering parameters from pionium data}

The results in this section for the positions and widths of the
pionium states of interest come from the formalism of Rasche and Woolcock
\cite{7}, which uses analytic continuation of the scattering matrix below
the threshold for a channel containing two oppositely charged hadrons.
These formal results have been confirmed by determining the position and
width directly, by studying the behaviour of the phase shift for the
(open) $\pi^{0}\pi^{0}$ channel. We also remark that the treatment of
Ref. \cite{7} is complete, and includes what are often (misleadingly)
called hadronic effects on the wavefunction at the origin. The results of
Ref. \cite{16} for the $ns$ states of pionium, which are quite different
from ours, are known to be wrong. New results for the $1s$ state are given
in a recent preprint by Sander and von Geramb \cite{19}, but no comment
is made on the dramatic difference between these new results and those of
Ref. \cite{16}. 

The width $\Gamma$ for the decay of pionium in the 1$s$ state to
$\pi^{0}\pi^{0}$ is given in Eqs. (2.10) and (2.13) of Ref. \cite{9} as 
\begin{equation}
\Gamma=\frac{8}{B}|1+p_{1}(1s)\varepsilon_{0}(1s)+\cdots|^{2}[-W_{C}
(1)]q_{o}^{}a_{oc}^{2}/(1+q_{o}^{2}a_{oo}^{2}),  \label{eq9}
\end{equation}
where 
\[
B=\left(\frac{1}{2}\alpha\mu_{c}\right)^{-1},  -W_{C}(1)=\frac{1}{4}
\alpha^{2}\mu_{c}, 
\]
\[
p_{1}(1s)=\frac{1}{2}(1+\gamma), {\rm Re}\varepsilon_{0}(1s)=4a_{cc}/B, 
\]
$\alpha$ being the fine structure constant and $\gamma$ Euler's constant.
The treatment of the $\pi^{-}p$ atom in Ref. \cite{7}, from which the
result, Eq. (\ref{eq9}), is taken, takes account neither of the relativistic
correction to the bound state position nor of the presence of the vacuum
polarisation potential in the calculation of $\Gamma$. The first is of no
consequence at the level of accuracy with which we are concerned. A full
treatment of the second has been worked out by Oades, Rasche and Woolcock
\cite{20} and leads to a small modification of Eq. (\ref{eq9}). The width in
the presence of vacuum polarisation is increased by the factor
$(1+3.08\times 10^{-3})$; this result has been obtained both analytically
and by a computer calculation of the width with and without the vacuum
polarisation potential. 

Strictly speaking, the elements of the scattering matrix appearing
in Eq. (\ref{eq9}) should be the elements of the K-matrix at the bound state
position, not at the $\pi^{+}\pi^{-}$ threshold. The only significant
effect is on $K_{oc}^{2}$; the small modification to Eq. (\ref{eq9}) is
estimated in Ref. \cite{9}  and amounts to a factor $(1-2.4\times 10^{-5})$. 

Eq. (\ref{eq9}) requires the values of $a_{cc}$ and $a_{oo}$, for which we
use the final numbers from Section 3, 
\[
a_{cc}=0.148\; {\rm fm},\; a_{oo}=0.053\; {\rm fm}. 
\]
With the other numbers 
\[
B  =  387.489\; {\rm fm}, \;\; -W_{C}(1)  =  1858.073\; {\rm eV},
\]
\[
q_{o}  =  35.509\; {\rm MeV}, \;\; p_{1}(1 s)  =  0.788608,
\]
\[
\hbar c  =  197.327053\; {\rm MeV}\; {\rm fm}, 
\]

we obtain the result 
\begin{equation}
a_{oc}({\rm fm})=0.3802[\Gamma({\rm eV})]^{\frac{1}{2}}.  \label{eq10}
\end{equation}
The uncertainty in the conversion constant of Eq. (\ref{eq10}) may be roughly
estimated as about 0.02$\%$. Most of it comes from the uncertainty in the
value of $a_{cc}$ used to calculate ${\rm Re}\varepsilon_{0}(1s)$. Since
the expected experimental accuracy in $\Gamma^{\frac{1}{2}}$ is about
5$\%$, the uncertainty in the conversion constant is of academic interest
only. 

The difference $\Delta W$ between the energies of the $2s$ and $2p$ 
levels has contributions from the relativistic
corrections, the vacuum polarisation potential and the presence of the
hadronic interaction. From the results of Austen and de Swart [21] the
relativistic corrections to the positions of the 2$s$ and 2$p$ states are 
\[
-37\mu_{c}\alpha^{4}/1024=-0.014\;{\rm eV}(2s),
\]
\[
-31\mu_{c}\alpha^{4}/3072=-0.004\; {\rm eV} (2p).
\]
Using a vacuum polarisation potential modified at short distances to take
account of the extended charge distributions of the pions, the
corresponding shifts in the positions of the 2$s$ and 2$p$ states are 
\[
-0.110\; {\rm eV}(2s), -0.004\;{\rm eV}(2p). 
\]
Thus 
\begin{equation}
\Delta W^{had}({\rm eV})=\Delta W({\rm eV})+0.116,  \label{eq11}
\end{equation}
where $\Delta W^{had}$ is the difference due to the hadronic interaction.

For the shifts in the positions of the 2$s$ and 2$p$ states due to
the hadronic interaction we again appeal to the results of Ref. \cite{7}.
For the 2$p$ state the shift is 
\[
\left(-\frac{1}{16}\alpha^{2}\mu_{c}\right)\left(\frac{3a_{cc}(\ell=1)}
{2B^{3}}\right)\approx 10^{-6}{\rm eV}, 
\]
which is completely negligible. Thus $\Delta W^{had}$ comes entirely from
the shift in the 2$s$ state, 
\begin{equation}
\Delta W^{had}=\left(-\frac{1}{16}\alpha^{2}\mu_{c}\right)\left(
\frac{2a_{cc}
} {B}\right)\left(1+p_{1}(2s)\frac{2a_{cc}}{B}+\cdots \right), \label{eq12}
\end{equation}
where 
\[
p_{1}(2s)={\rm ln}2+\gamma-\frac{1}{2}=0.770363. 
\]
It follows from Eq. (\ref{eq12}) that 
\begin{equation}
a_{cc}({\rm fm})=-0.4168\Delta W^{had}({\rm eV}).  \label{eq13}
\end{equation}
From our present knowledge of $a_{cc}$ we expect that $\Delta W^{had}$ 
will be close to $-0.355$ eV. The experimenters therefore face the
difficult task of measuring with reasonable accuracy a difference
$\Delta W$ which is about $-0.47$ eV, compared with the position of the
$n$=2 pure Coulomb levels 464.52 eV below the $\pi^{+}\pi^{-}$ threshold.

\section{Electromagnetic corrections to the threshold scattering matrix}

The final step is to calculate the difference ${\bf {\tilde{a}}-{a}}$
, where ${\bf {\tilde{a}}}$ and ${\bf {a}}$ are defined in Eqs. (\ref{eq3}) 
and (\ref{eq7}) respectively. As explained in Section 1, the first step is
to construct energy independent double square well potentials, using the
Schr\"{o}dinger equation \ref{eq5}, which reproduce as accurately as
possible up to 500 MeV the phase shifts $\delta_{0}^{0}$ and
$\delta_{0}^{2}$ provided by Gasser \cite{13}.
There are inner and outer ranges of the potentials to be varied, and it
was found that the best fits are obtained with ranges 
\[
r_{1}=0.3\;{\rm fm},\; r_{2}=1.5\; {\rm fm}. 
\]
The full potentials which we used to obtain the final results are, to 4
significant figures, 
\begin{eqnarray}
V_{0}^{0}(r)  & = & -6824\; {\rm MeV}, \;\; 0 \leq r < 0.3\; {\rm fm},
\nonumber \\
& = & +102.0\; {\rm MeV}, \;\; 0.3\; {\rm fm} \leq r < 1.5\; {\rm fm},
\nonumber \\
& = & 0, \;\; r \geq 1.5\; {\rm fm};  \label{eq14}
\end{eqnarray}
\begin{eqnarray}
V_{0}^{2}(r) & = & +39980\; {\rm MeV}, \;\; 0 \leq r < 0.3\; {\rm fm},
\nonumber \\
& = & -53.57\; {\rm MeV}, \;\;0.3\; {\rm fm}  \leq r <1.5\; {\rm fm},
\nonumber \\
& = & 0, \;\; r \geq 1.5\; {\rm fm}.  \label{eq15}
\end{eqnarray}

The corresponding scattering lengths, which agree with those
obtained by extrapolating the input phase shifts to $W=2\mu_{c}$, are 
\begin{equation}
a_{0}^{0}=0.2883\; {\rm fm}\,,\; a_{0}^{2}=-0.0617\; {\rm fm}\,,
\label{eq16}
\end{equation}
and the matrix ${\bf {\tilde{a}}}$ of Eq. (\ref{eq7}) is 
\begin{equation}
{\bf {\tilde{a}}=\left(
\begin{array}{cc}
0.1716\; {\rm fm}\, & -0.1650 \;{\rm fm}\, \\ 
-0.1650\; {\rm fm}\, & 0.0550 \;{\rm fm}\,
\end{array}
\right).}  \label{eq17}
\end{equation}

Fig. 1 shows that the double square well potentials of Eqs. (\ref{eq14})
and (\ref{eq15}) reproduce the phase shifts of Ref. \cite{13} remarkably well.
Not surprisingly, this agreement deteriorates above 500 MeV but this is
of no concern. The simple effective potentials clearly represent the
hadronic interaction very well over an energy range which is quite
sufficient for our purpose. 

The scattering lengths (Eq. (\ref{eq16})) come predominantly from the inner
parts of the potentials in Eqs. (\ref{eq14}) and (\ref{eq15}) and the
inner radius of 0.3 fm is approximately the smallest radius used by
Moor et al. \cite{9} for their single square well potentials. The use of
the much larger outer radius of 1.5 fm with the small outer potential,
is a simple way of getting good fits to the phase shifts. The
scattering lengths in Eq. (\ref{eq16}) differ from those of Ref. \cite{9}
mainly because the reference mass used in the $\chi$PT equations is
$\mu_{c}$, as explained in Section 1. 

The Coulomb potential $V^{C}(r)$ for extended charge distributions
was taken as the potential between uniformly charged spheres, exactly as
in Ref. \cite{9} (last paragraph of Section 4). We have checked that using
gaussian charge distributions instead makes no difference to the
scattering matrix in Eq. (\ref{eq18}) at the level of accuracy given. 

With the input potentials ${V}^{C}(r)$ just described and ${\bf
{V^{had}}}$ obtained from Eqs. (\ref{eq5}), (\ref{eq14}) and (\ref{eq15}),
the threshold scattering matrix arising from Eq. (\ref{eq8}) is 
\begin{equation}
{\bf {a}=\left(
\begin{array}{cc}
0.1479\; {\rm fm}\, & -0.1599 \;{\rm fm}\, \\ 
-0.1599 \;{\rm fm}\, & 0.0533 \;{\rm fm}\,
\end{array}
\right).}  \label{eq18}
\end{equation}
The desired electromagnetic corrections are therefore 
\begin{equation}
\tilde{a}_{cc}-a_{cc}=0.0237\;{\rm fm}\,,\;\tilde{a}_{oc}-a_{oc}=-0.0051\; 
{\rm fm},  \label{eq19}
\end{equation}
or, with the definitions 
\[
\tilde{a}_{cc}=(1+\Delta_{cc})a_{cc},\; \tilde{a}_{oc}=(1+
\Delta_{oc})a_{oc}, 
\]
\begin{equation}
\Delta_{cc}=0.160,\; \Delta_{oc}=0.032.  \label{eq20}
\end{equation}
It is of interest to note that the contribution to the corrections from the
pion mass difference is far greater than that from the Coulomb potential.
The detailed numerical results are given in Table 1.

On the basis of many calculations which we have made using different
scattering lengths $a_{0}^{0}$ and $a_{0}^{2}$ as the starting values, we
have found that the relative correction $\Delta_{oc}$ is much less
sensitive to these starting values than is the difference 
$\tilde{a}_{oc}-a_{oc}$. However, the difference $\tilde{a}_{cc}-a_{cc}$
is less sensitive than $\Delta_{cc}$. The uncertainties in the values of
$\tilde{a}_{cc}-a_{cc}$ and 
$\Delta_{oc}$ need to be estimated. Even though some fine details of the
potentials have no doubt been missed, we are confident that the use of
double square well potentials introduces a negligible error. We have
estimated the uncertainties due to the variation of the hadronic potentials
within limits which correspond to the uncertainties in the input phase
shifts and in our fitting procedure. This effectively includes the
uncertainties in the input scattering lengths. The final results are 
\begin{equation}
\tilde{a}_{cc}-a_{cc}=0.024\pm 0.002 \;{\rm fm},\; \Delta_{oc}=0.032 \pm
0.002,  \label{eq21}
\end{equation}
where we have kept the number of significant digits appropriate to the size
of the estimated errors. The development of two-loop $\chi$PT \cite{22}
will result in slightly different phase shifts, but the resulting changes
in $\tilde{a}_{cc}- a_{cc}$ and in $\Delta_{oc}$ will lie within the
uncertainities given in Eq. (\ref{eq21}).  

The values of $a_{oc}$ and $a_{cc}$ obtained from experimental data
on pionium must be corrected using Eq. (\ref{eq21}) , and the resulting
values of $\tilde{a}_{oc}$ and $\tilde{a}_{cc}$ compared with the values
of $a_{0}^{0}$ and $a_{0}^{2}$ obtained from a combination of $\pi\pi$
scattering data at energies above $W=2\mu_{c}$ and the constraints of
$\chi$PT. If this comparison should eventually reveal a significant
discrepancy, it will then be necessary to look again at what is known
about $\pi\pi$ scattering above the $\pi^{+}\pi^{-}$ threshold. 

\acknowledgements
We thank L.L. Nemenov for useful discussions and J. Gasser for providing us
with the phases predicted by single-loop $\chi$PT. We also thank the Swiss
National Foundation for financial support.

\newpage
\begin{figure}
\caption{The $I=0$ and $I=2$ $\pi\pi$ s-wave phase shifts $\delta_{0}^{0}$
and $\delta_{0}^{2}$ given by $\chi$PT
(Ref. \protect\cite{13}) and by the double square wells of Eqs. 
(\protect\ref{eq14}) and (\protect\ref{eq15}).}
\end{figure}

\begin{table}
\caption{The separate effects of the pion mass difference and of
the Coulomb potential on the elements of the scattering matrix at the
$\pi^{+}\pi^{-}$ threshold (all results in fm) }
\begin{tabular}{llll}
\hline
& \multicolumn{1}{c}{$a_{cc}$} & \multicolumn{1}{c}{$a_{oc}$} & 
\multicolumn{1}{c}{$a_{oo}$} \\ \hline
hadronic & 0.1716 & $-0.1650$ & 0.0550 \\ 
with Coulomb potential only & 0.1719 & $-0.1662$ & 0.0561 \\ 
with mass difference only & 0.1478 & $-0.1589$ & 0.0522 \\ 
fully corrected & 0.1479 & $-0.1599$ & 0.0533 \\ \hline
\end{tabular}
\end{table}\

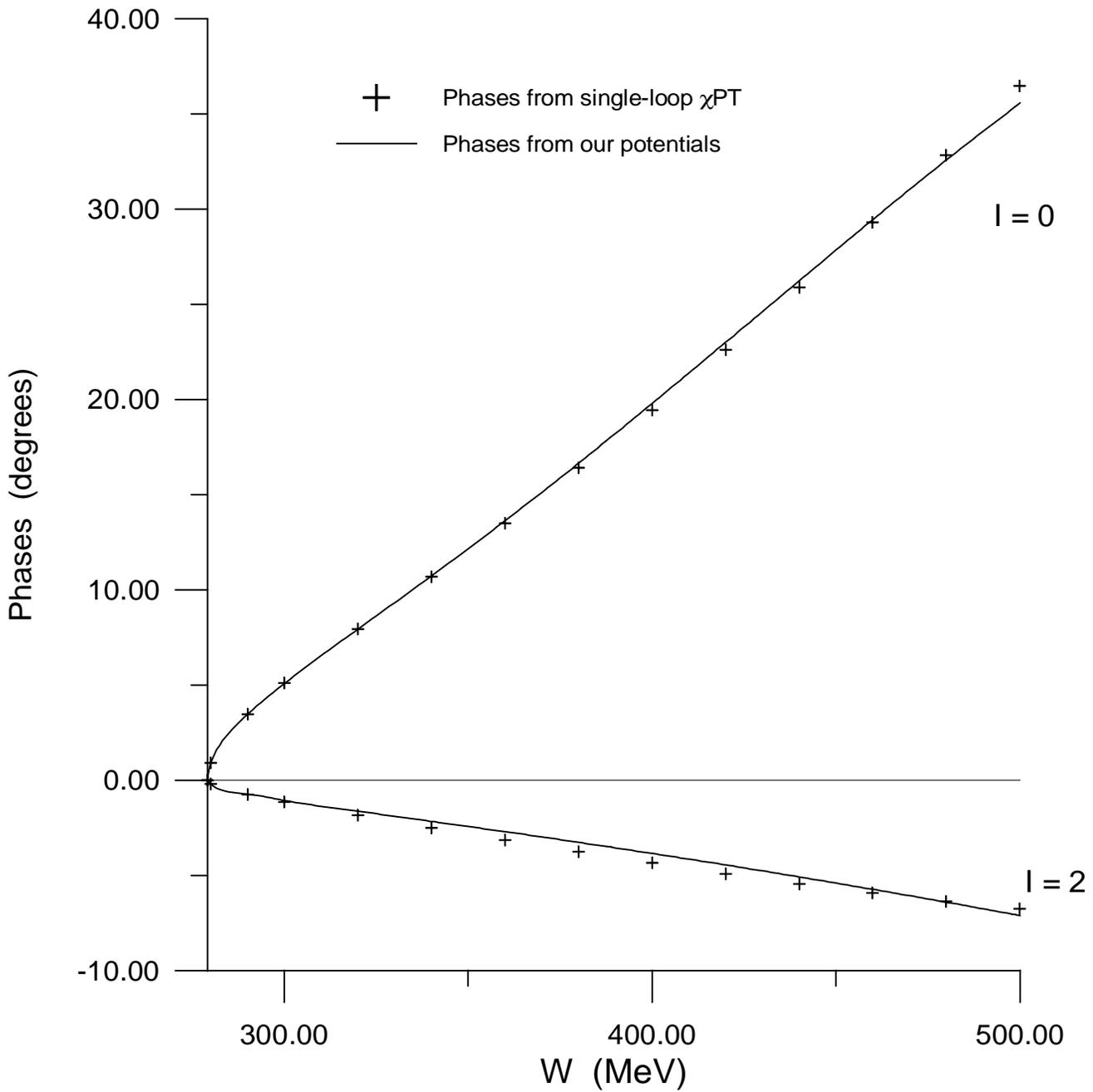
\end{document}